\documentclass[a4paper]{jpconf}
\usepackage{amsmath}
\usepackage{amssymb}
\usepackage{graphicx}
\usepackage{amsfonts}
\usepackage{bm}
\newcommand{\be}{\begin{equation}}
\newcommand{\ee}{\end{equation}}
\newcommand{\bea}{\begin{eqnarray}}
\newcommand{\eea}{\end{eqnarray}}
\newcommand{\hf}{\frac12}
\newcommand{\nn}{\nonumber\\}
\def\journal#1#2#3#4{#4 {\it #1} {\bf #2} #3}

\def\ih{\frac{i}\hbar}
\def\la{\langle}
\def\ra{\rangle}

\def\eq#1{(\ref{#1})}
\def\ord#1{{\cal O}(#1)}

\def\Tr{{\mathrm{Tr}}}

\def\hj{\hat j}
\def\hx{\hat x}
\def\hD{\hat D}

\begin{document}
\title{Environment Induced Time Arrow and the Closed Time Path method}
\author{Janos Polonyi}
\address{Strasbourg University, High Energy Theory Group, CNRS-IPHC,
23 rue du Loess, BP28 67037 Strasbourg Cedex 2 France}
\ead{polonyi@iphc.cnrs.fr}
\begin{abstract}
It is shown in the framework of a harmonic system that the thermodynamical time arrow is induced by the environmental initial conditions in a manner similar to spontaneous symmetry breaking. The Closed Time Path formalism is introduced in classical mechanics to handle Green functions for initial condition problems by the action principle, in a systematic manner. The application of this scheme for quantum systems shows the common dynamical origin of the thermodynamical and the quantum time arrows. It is furthermore conjectured that the quantum-classical transition is strongly coupled.
\end{abstract}

\section{Introduction}
The fundamental equations of motions are time reversal invariant when weak interaction is ignored and the emergence of past and future, an orientation of time is an interesting, open question \cite{zehta,savitt,mackey}. One usually distinguishes radiational (use of retarded Green functions in classical electrodynamics), thermodynamical (second law of thermodynamics), quantum (information loss at the quantum-classical crossover) and gravitational (singularity generated time reversal asymmetry) time arrows. It is argued below that the thermodynamical time arrow, irreversibility, is generated by a spontaneous breakdown of time reversal invariance and has the same dynamical origin as decoherence, thereby the thermodynamical and quantum time arrows are identical. These points will be demonstrated in the framework of harmonic systems. To render this argument valid for weakly interactive systems one needs the Closed Time Path (CTP) formalism which will briefly be introduced in classical mechanics \cite{arrow}. 

The construction presented here is part of a long time project, the derivation of classical electrodynamics from QED \cite{ed}. Due to the large number of particles involved, the description of the quantum-classical crossover must be based on quantum field theory. The best way to cross the quantum-classical transition is to follow the renormalization group trajectory which has to be set up for the density matrix, rather than transition amplitudes. This is the point where the CTP formalism appears in the discussion. It is not difficult to find the CTP renormalization group equation \cite{ed}, however its solution, more precisely the construction of a reliable, treatable approximate solution requires the clarification of several details. One needs a handle on the dynamical build up of decoherence \cite{coulomb}, the equation of motion for the expectation value and the decoherence of composite operators for free particles \cite{nv}, the quantum corrections to the equation of motion of elementary operators \cite{maxwell}, the characterization of the classicality of a scattering process \cite{scattenv} to mention some of them. A further issue, the relation between irreversibility and decoherence \cite{irrac,arrow} is discussed briefly here.

The point where the quantum field theoretical description becomes important is the generation of new scales by the help of the number of particles involved, $N\sim10^{23}$. The probability distributions become very narrow, $\Delta p/p\sim10^{-12}$ and the central limit theorem extended by perturbation expansion may produce deterministic laws for its peak around an IR fix point. It is no longer necessary to assume the collapse of the wave function, $N$ may generate short time scales and the state reduction can take place in extremely short time, requiring the use of relativistic quantum field theory. The way a charged particle is detected in the Wilson cloud chamber suggests that the number of particles, participating in a measurement may depend strongly on time, $N_{eff}(t)$ and can produce highly non-linear effects.

The presentation starts by pointing to the initial conditions for the environment, as the origin of the thermodynamical time in Section \ref{tarrows} and continues with a simple, classical harmonic system displaying irreversibility in Section \ref{clharms}. The view of irreversibility as a dynamical symmetry breakdown is motivated in Section \ref{irrs}. In Section \ref{cctps} the CTP formalism is introduced in classical mechanics, quantum systems are addressed in Section \ref{qsysts}. Finally, Section \ref{summs} contains a brief summary.

\section{Thermodynamical time arrows}\label{tarrows}
The thermodynamical time arrow, irreversibility, is easiest to understand in terms of coarse graining, or blocking in the language of renormalization group. Let us consider a classical dynamical system, described by the coordinates $x$ and $y$, called system and its environment. The former component comprises the degrees of freedom observed. The complete trajectory is supposed to obey a set of reversible equations of motion, 
\be
\ddot x=F(x,y),~~~~~~\ddot y=G(x,y).
\ee
The time arrow is inferred for each degree of freedom by observation: One introduces an external perturbation of the degree of freedom in question which acts at a given time $t_p$ and if the response is seen for $t>t_p$ or $t<t_p$ then the experimental time arrow is forward or backward, respectively. 

Such a a time arrow can be set either in a trivial, kinematical or a highly non-trivial, dynamical manner. The kinematical way is the imposition of some boundary conditions in time which renders the solution unique. In fact, by imposing initial or final conditions, independently on the external perturbations, on a certain degree of freedom we construct its motion forward or backward in time. It may happen that the kinematical time arrows, introduced in this manner for several degrees of freedom are in conflict and the complete system has no causal behavior and well defined time arrow. 

A time arrow is set dynamically and the corresponding dynamics is irreversible if the time arrow is formed even before imposing the boundary conditions on time on the system coordinate. This may happen if the environment time arrows, as external agents manage to break the time reversal invariance of the effective system dynamics. The setting of a kinematical system time arrow, identical to the environment leads to the usual irreversible dynamics but the opposite kinematical time arrow generates acausal effective system dynamics, without well defined system time arrow. It is assumed in what follows that the kinematical time arrow is identical, forward pointing, for all degree of freedom.

The system dynamics is constructed by eliminating the environment by means of its equation of motion, $y\to y[x;y(0),\dot y(0)]$, $y(0)$ and $\dot y(0)$ denoting the environment initial conditions, imposed at the initial time $t_i=0$. The insertion of this solution into the system equation of motion yields the effective system dynamics,
\be\label{effem}
\ddot x=F_{eff}[x;y(0),\dot y(0)].
\ee
Time reversal of the trajectories for the time interval $0<t<t_f$ acts as $Tx(t)=x(t_f-t)$, it exchanges initial conditions into final conditions, $Ty(0)=y(t_f)$ and the time reversed effective equation of motion \eq{effem} can be written as
\be
T\ddot x=F_{eff}[Tx;y(t_f),\dot y(t_f)].
\ee
Since $y(t_f)\ne y(0)$ and $\dot y(t_f)\ne\dot y(0)$ the thermodynamical system time arrow is set by the environment initial conditions. 

It will be found below that a necessary condition for irreversibility is a condensation point in the normal mode spectrum and the emergence of the thermodynamical time arrow may lead to acausality, the response to an external effect may show up before the effect. This latter phenomenon is rather puzzling because causality, the precedence of an external perturbation of its response, seems to be granted automatically when Newton equations  are integrated in time.

\section{Classical harmonic systems}\label{clharms}
Let us consider a harmonic dynamics, governed by the Lagrangian
\be\label{clho}
L=\frac{m}{2}\dot x^2-\frac{m\omega_0^2}2x^2+jx+\sum_n\left(\frac{m}2\dot y^2_n-\frac{m\omega_n^2}2y^2_n+g_nxy_n\right),
\ee
where the condition $\sum_ng_n^2/m\omega^2_n<m\omega_0^2$ guarantees the stability of the motion. We impose the environment initial conditions $y_n(0)=\dot y_n(0)=0$ and write the system trajectory in the form
\be\label{greenvsol}
x(t)=-\int_0^\infty dt'D^r(t-t')j(t')
\ee
where the retarded Green function,
\be
D^r(\omega)=\frac1{m[(\omega+i\epsilon)^2-\omega^2_0]-\Sigma^r(\omega)},
\ee
is given in terms of the retarded self energy
\be
\Sigma^r(\omega)=\sum_n\frac{g^2_n}{m(\omega+i\epsilon)^2}.
\ee
The sign of $\epsilon$ reflects the environment time arrow, one needs $\epsilon=0^+$ or $\epsilon=0^-$ for initial or final environment conditions. We use forward environment time arrow below, $\epsilon=0^+$. The spectrum, defined by the equation $0=D^{r-1}(\omega_j+i\epsilon)$ consists of frequencies with negative, $\ord{\epsilon}$ imaginary part, hence the effective system dynamics is reversible (infinite life-time) and causal (poles below the real axis).

It is advantageous to introduce the spectral function
\be
\rho(\Omega)=\sum_n\frac{g_n^2}{2m\omega_n}\delta(\omega_n-\Omega)
\ee
whose phenomenological, Ohmic form, \be
\rho(\Omega)=\frac{\Theta(\Omega)g^2\Omega}{m\Omega_D(\Omega_D^2+\Omega^2)}
\ee
leads to
\bea\label{oselfen}
\Sigma^r(\omega)&=&\frac{g^2}{m\Omega_D}\int_0^\infty d\Omega\frac{\Omega}
{\Omega_D^2+\Omega^2}\frac{2\Omega}{(\omega+i\epsilon)^2-\Omega^2}\nn
&=&-\frac{i\pi g^2}{\Omega_D(\omega+i\Omega_D)}
\eea
and
\be
D^r(\omega)=\frac1{m[(\omega+i\epsilon)^2-\omega^2_0]+\frac{i\pi g^2}{\Omega_D(\omega+i\Omega_D)}}.
\ee
The lesson of this Green function is that in case of continuous spectrum, in particular for the Ohmic spectral function the effective dynamics is irreversible (finite life-time) and acausal (poles above the real axis).

\section{Irreversibility}\label{irrs}
The goal of this section is to underline the similarities between spontaneous symmetry breaking and irreversibility. There are two ways to detect spontaneous symmetry breaking, a dynamical and a static procedure.

\subsection{Slow order parameter} 
We see a finite part of our Universe which requires a more refined understanding of spontaneous symmetry breaking in large but finite systems. Spontaneous symmetry breaking can be understood dynamically as the slowing down of the order parameter. In fact, consider the rotation of a macroscopic, rigid body consisting of a large number, $N\sim10^{23}$, of particles and ignore first its environment. The rigid rotation is driven by the Hamiltonian
\be
H_{rot}=\hf L^j(\Theta^{-1})^{jk}L^k+\ord{L^4}
\ee
where the inertia tensor is $\Theta^{jk}=m\sum_nx^j_nx^k_n=\ord{N}$. The ground state is non-degenerate but a macroscopic systems cannot be isolated from their environment hence we have to assume $L\sim n\hbar$ where $n$ is a not a too large number. The angular velocity, $\omega^j=(\Theta^{-1})^{jk}L^k=\ord{N^{-1}}$, is extremely slow, a full rotation takes approximately $10^{10}/n$ times the age of our Universe. Therefore, the spontaneous breakdown of rotational symmetry is the applicability of the adiabatic approximation for the slow order parameters, the Euler angles in this case.

Such a slowing down appears as a violation of Fubini theorem in perturbation expansion, the non-commutativity of the thermodynamical limit, $N\to\infty$ and the limit of long observation time, $T\to\infty$,
\be
\lim_{N\to\infty}\lim_{T\to\infty}A(N,T)\ne
\lim_{T\to\infty}\lim_{N\to\infty}A(N,T).
\ee
The discrete spectral lines of a large but finite system are resolved in the first case but the infinite system, observed in finite time allows us the use of continuous spectrum in the second order of limits and the scenario described in Section \ref{clharms} is recovered.

A mathematical ambiguities, such as the non-commutativity of limits always correspond to a physical alternative, the possibility of different preparation or observation of the system. Irreversibility is detected by observations which are limited in time and which, as a result, can not resolve the discrete spectrum. This happens if the spectral lines have an unexpectedly high density. This is reminiscent of quantum anomalies, the finite, observable UV cutoff effects of renormalizable quantum field theories due to the non-uniform convergence of the bare, regulated Feynman loop-integrals, except that irreversibility arises due to the smearing out an IR cutoff effect, the discreteness of spectral lines. In this sense spontaneous symmetry breaking, in particular irreversibility is an IR anomaly.

\subsection{Observations with IR cutoff}
\begin{figure}
\begin{center}
\includegraphics[width=4.5cm]{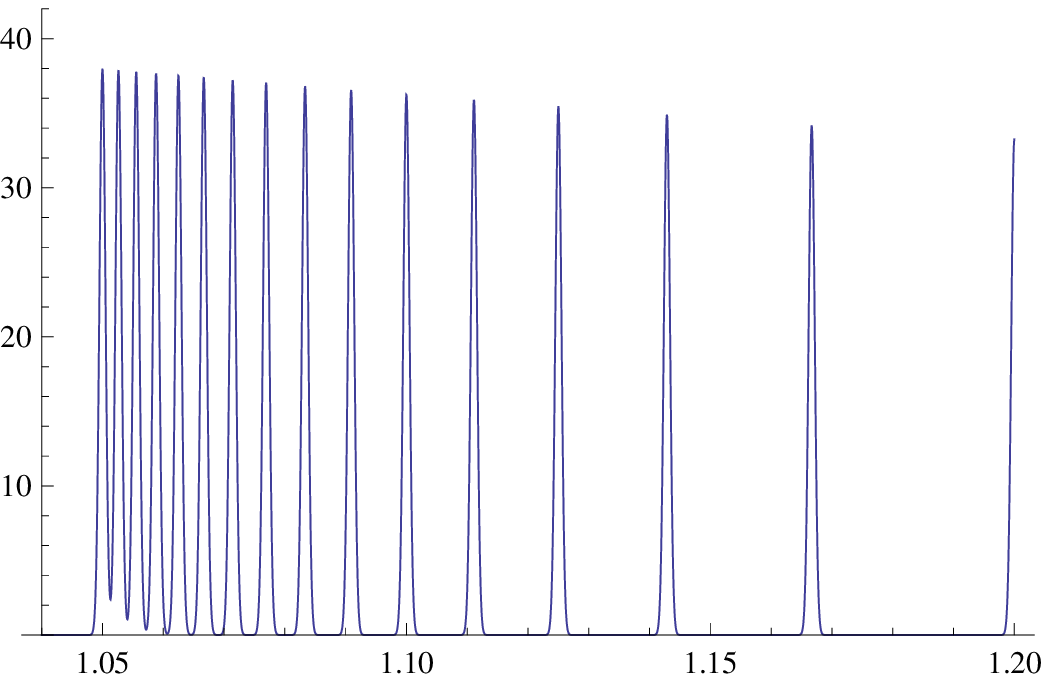}
\includegraphics[width=4.5cm]{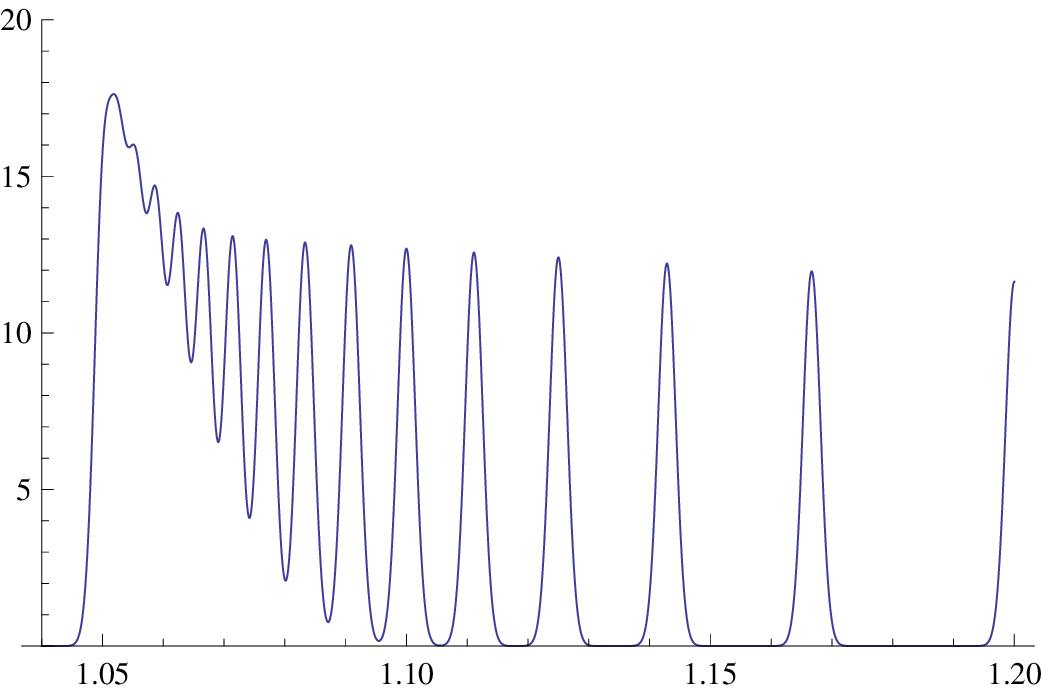}
\includegraphics[width=4.5cm]{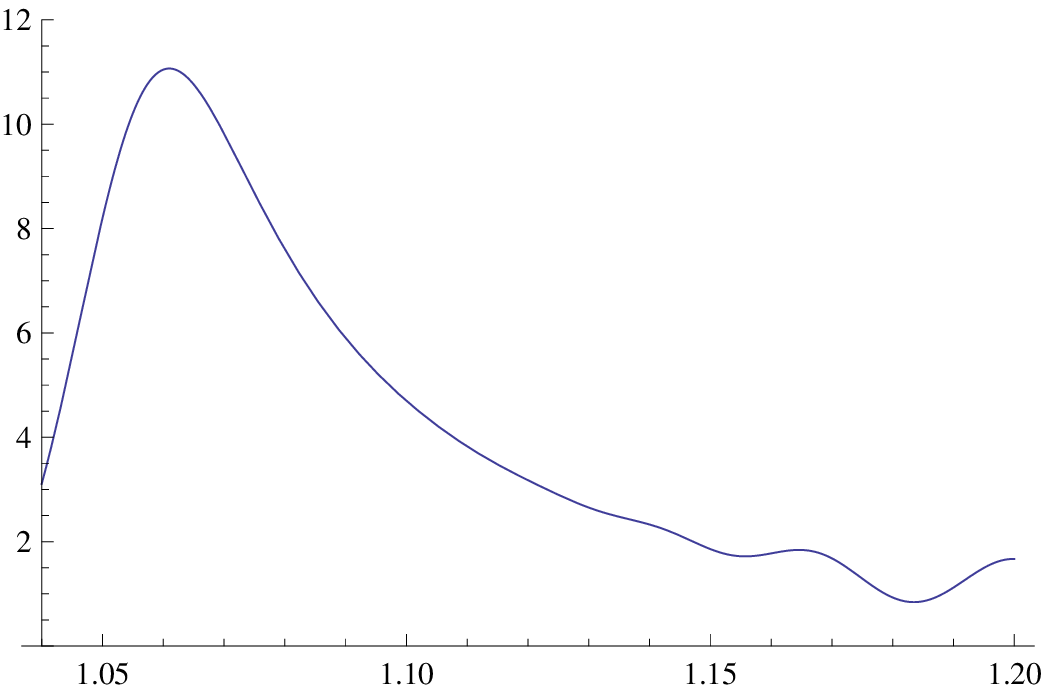}

(a)\hskip4.5cm (b)\hskip4.5cm (c)
\end{center}
\caption{The apparent spectral density, \eq{appspfc} for observation times (a): $T=2000$, (b) $T=700$ and (c) $T=100$. The spectrum $\omega_j=1+1/j$ is constructed by $N=20$, $a_{kj}=1/\sqrt{N}$, $m=1$  and the IR cutoff is $c(t)=\exp(-t^2/2T^2)$.}\label{splines}
\end{figure}

It is instructive to follow what happens in a simple harmonic system given by the coordinates $x_n$ if the limitation on observation time is taken into account by writing an observed normal mode $\tilde x_j=\sum_na^{-1}_{jn}x_n$ as $x_{obs}(t)=c(t)x(t)$ where the function $c(t)$, satisfying the conditions $c(0)=1$, $\lim_{t\to\infty}c(t)=0$ acts as an IR cutoff. It is well known that the finite observation time spreads the discrete spectral lines of the harmonic system. In fact, an external source $j(t)=j_k\delta(t)$, coupled linearly to normal mode $\tilde x_k$ yields the response
\be\label{obsnm}
\tilde x_{k~obs}(\omega)=-j_k\int\frac{d\Omega\rho_m(\Omega)}{\omega+i\epsilon-\Omega},
\ee
where the spectrum weight,
\be
\rho_m(\Omega)=\sum_{j=1}^N\delta(\Omega-\tilde\omega_j)\frac{a^2_{kj}}{2m\tilde\omega_j},
\ee
inferred by infinitely long observations for a discrete system is replaced by the apparent spectral function
\be\label{appspfc}
\rho_m(\Omega)=\frac1{2\pi}\sum_{j=1}^Nc(\Omega-\tilde\omega_j)\frac{a^2_{kj}}{2m\tilde\omega_j}
\ee
in the integral of Eq. \eq{obsnm}. The apparent spectral function, \eq{appspfc}, depicted in Fig. \ref{splines} shows that if a discrete spectrum has a condensation point then arbitrarily long, but finite time observations leave infinitely many spectrum lines unresolved. The dense spectral lines of the unresolved normal modes make the continuous spectrum approximation applicable at the condensation point and these normal modes act as a sink for energy, generating dissipation and irreversibility if they have sufficient spectral weight. If the spectrum is uniformly dense in the thermodynamical limit, as in field theories then dissipation to the unresolved normal modes and irreversibility takes place in the form of Bremsstrahlung at any energy.

Acausality can be understood in a similar manner, as a result of uncertainties in determining the time when the external source $j(t)$ acts on the system. The relation
\be
x^{obs}(\omega)=-\int\frac{d\omega'}{2\pi}c(\omega-\omega')D^r(\omega')j(\omega')
\ee
between the external source and the limited observation allows us to infer an apparent source
\bea
j^{obs}(\omega)&=&-D^{r-1}(\omega)x^{obs}(\omega)\nn
&=&\int\frac{d\omega'}{2\pi}c(\omega-\omega')\frac{D^r(\omega')}{D^r(\omega)}j(\omega'),
\eea
The phenomenological form: $D^r(\omega)=\sum_pZ_p/(\omega-\omega_p)$ gives the pole contribution
\be
j^{obs}_p(\omega)\approx\int\frac{d\omega'}{2\pi}c(\omega-\omega')\frac{\omega-\omega_p}{\omega'-\omega_p}j(\omega')
\ee
for $\omega\sim\omega_p$. The IR cutoff $c(\omega)=2\eta/(\omega^2+\eta^2)$, with $\eta=2\pi/T$ and the source: $j(t)=j_0\delta(t)$ yields the contribution
\be
j_p^{obs}(t)=j_0\left[\delta(t)-\frac{2\pi}Te^{-i\omega_pt-2\pi\frac{|t|}T}\right]
\ee
to the apparent source. Hence the apparent acausality results from difficulties in reconstructing a sharply localized external source in time due to the abundant soft modes. When a finite time step $\Delta t$ is used to integrate Newton equations then acausality appears as an ambiguity in the order of the steps $N\to\infty$ and $t\to t+\Delta t$.

\subsection{Weak explicit symmetry breaking} 
Spontaneous symmetry breaking can be detected in a static, equilibrium state, too: One introduces an external symmetry breaking and lets its strength tend to zero. Spontaneous symmetry breaking takes place if the system does not "forget" the external symmetry breaking in this limit. The analogue of external symmetry breaking in case of irreversibility is the environment boundary condition. Thus it is illuminating to follow the impact of environment time arrow in the effective system dynamics. 

Let us consider the response of the system coordinate to a source $j(t)$, either an environment coordinate or an external diagnostic tool, given in the form of Eq. \eq{greenvsol}. The symmetric or antisymmetric part of the retarded Green function with respect to time reversal is the near or far Green function, respectively, hence we have $D^r=D^n+D^f$, with $D^n(t)=D^n(-t)$ and $D^f(t)=-D^f(-t)$. In case of a harmonic oscillator of mass $m$ and frequency $\omega_0$ we have
\be
D^n(\omega)=P\frac1{m(\omega^2-\omega_0^2)},~~~~~~
D^f(\omega)=-\frac{\pi i}m\mathrm{sign}(\omega)\delta(\omega^2-\omega_0^2),
\ee
where $P$ denotes the principal part. The support of $D^f$ is the mass-shell, the null-space $D^{-1}x=0$ of the linear equation of motion, $D^{-1}x=j$, where $D^n$ is vanishing. Therefore the time reversal symmetric response to $j(t)$ is generated by the off-mass shell modes of  $D^n$ and the time arrow is set by the mass shell modes of $D^f$. Such a transmutation of the negative time parity couplings can be seen in the self energy \eq{oselfen}. In fact, the $i\epsilon$ term in the environment Green function can be viewed as a weak, external time reversal invariance breaking because it changes sign when the environment initial conditions are replaced by final conditions. The response of the infinitesimal $i\epsilon$ term is the finite imaginary part of the system self energy, $\Im\Sigma^r(\omega)$, which changes sign under time reversal, too.

\section{Classical Closed Time Path formalism}\label{cctps}
The CTP formalism has quantum origin, it has been introduced first by J. Schwinger to establish perturbation expansion in the Heisenberg representation \cite{schw}. It has been reinvented in several other contexts such as to derive relaxation in many-body systems \cite{kadanoffb}, to find the mixed state contributions to the density matrix by path integral \cite{feynmanv,diosi}, to develop perturbation expansion for retarded Green-functions \cite{keldysh}, to find manifestly time reversal invariant description of quantum mechanics \cite{aharonov}, to describe finite temperature effects in quantum field theory \cite{umezawa,niemisa,niemisb}, to formulate the consistency of histories \cite{griffith,omnes,gellmann}, to derive equations of motion for the expectation value of local operators \cite{jordan,ed}, to follow non-equilibrium processes \cite{calzetta} and to describe scattering processes with non-equilibrium final states \cite{scattenv}. This scheme will be introduced below in classical mechanics to allow us to recast initial problems as variational equations and to facilitate the introduction of retarded Green functions. The distinguishing feature of CTP is the redoubling of the degrees of freedom, coming from the double occurrence of the time evolution operator $U(t)=\exp(-iHt/\hbar)$ in the Heisenberg representation, $A(t)=U^\dagger(t)A(0)U(t)$. Apart of this physical origin, the doubling facilitates the handling of dissipative forces by variation principle, as well \cite{bateman,vitiello,arrow}.

\subsection{Action}
The variational principle can not be used for initial condition problems because the variation at the final time cancels the momentum,  $\partial S/\partial x_f=p(t_f)=0$. The remedy of this problem is the flipping the time arrow at $t_f$ and the sending of the whole system which obeys reversible dynamics back to its initial conditions. We impose the initial condition $x(0)=\dot x(0)=0$ and use the corresponding trajectory $\tilde x(t)$, $0<t<2t_f$ to construct a CTP doublet, $\hx(t)=(x^+(t),x^-(t)$ with $x^+(t)=\tilde x(t)$ and $x^-(t)=\tilde x(2t_f-t)$. The action for the CTP doublet trajectory is
\be\label{ctpaction}
S_{CTP}[\hx]=\int_0^{t_f}dt[L_{CTP}(x^+(t),\dot x^+(t))-L_{CTP}^*(x^-(t),\dot x^-(t))]
\ee
where the minus sign and the complex conjugation are due to the flipping of the time arrow.The Lagrangian 
\be
L_{CTP}(x,\dot x)=L(x,\dot x)+i\frac{m\epsilon}2x^2
\ee
contains an infinitesimal imaginary part to remove the degeneracy of the action for $x^+=x^-$. Note that the condition $x^+(t_f)=x^-(t_f)$ closes the trajectories and renders the variational equation trivial, $p(t_f)-p(t_f)=0$, at the final time. The possibility of irreversible effective system dynamics, signaled by $x^-(t)\ne x^+(t)$ arises because the environment time arrow is not flipped at the final time and the system and environment time arrows end up in conflict.

There is an important symmetry in the CTP formalism, the exchange of the trajectories, $\tau x^\pm=x^\mp$, in particular $\tau S_{CTP}[\hx]=-S^*_{CTP}[\hx]$.

\subsection{Green function}
The action of a harmonic system,
\be
S_{CTP}[\hat\phi]=\int dx\left[\hf\hat\phi(x)\hat K\hat\phi(x)+\hj(x)\hat\phi(x)\right],
\ee
identifies the CTP Green function $\hD=\hat K^{-1}$. The CTP symmetry $\tau$ requires the form
\be\label{qctpa}
\hat K=\begin{pmatrix}K^n+i\bar K_1&K^f-i\bar K_2\cr-K^f-i\bar K_2&-K^n+i\bar K_1\end{pmatrix},
\ee
which results in the parametrization
\be
\hD=\begin{pmatrix}D^n+i\bar D_1&-D^f+i\bar D_2\cr D^f+i\bar D_2&-D^n+i\bar D_1\end{pmatrix},
\ee
of the Green function in terms of four real functions, $\hat K^{tr}=\hat K$, $\hD^{tr}=\hD$, $\bar D_1^{tr}=\bar D_1$, $\bar D_2^{tr}=\bar D_2$, $(D^n)^{tr}=D^n$, $(D^f)^{tr}=-D^f$. A physical source, $j^\pm(t)=\pm j(t)$ generates the trajectory 
\be\label{phctptr}
x^\sigma(t)=-\sum_{\sigma'=\pm1}\int_0^{t_f}dt'D^{\sigma\sigma'}(t-t')\sigma'j(t')
\ee
and the condition $x^+=x^-$, expressed as
\be\label{ctpcons}
D^{++}-D^{+-}=D^{-+}-D^{--},
\ee
reduces the number of functions in the Green function to three, $\bar D_1=\bar D_1=\bar D$. It is easy to see that the retarded and advance Green functions are defined by $D^{\stackrel{r}{a}}=D^n\pm D^f$. Such an origin of the Green function explains that $\bar D$ drops out from Eq. \eq{phctptr} and remains arbitrary in classical dynamics. Nevertheless $\bar D$ plays an important role in quantum dynamics as we shall see below.

\section{Quantum systems}\label{qsysts}
The CTP formalism is needed in quantum physics when the density matrix $\rho(t)=U(t,0)\rho(0)U^\dagger(t,0)$ or expectation values, $\la A\ra(t)=\Tr[A \rho(t)]$ are sought. The reduplication of the degrees of freedom is due to the need of the independent construction of the time evolution operator $U$ and $U^\dagger$, achieved by means of the generator functional
\be\label{ctpgf}
e^{iW[\hj]}=\Tr T[e^{-\ih\int dt'[H(t')-j^+(t')x(t')]}]\rho_iT^*e^{\ih\int dt'[H(t')+j^-(t')x(t')]}],
\ee
where $T^*$ is anti-time ordering. The path integral representation,
\be\label{ctpgenvfpi}
e^{iW[\hj]}=\int D[\hx]e^{\ih S_{CTP}[\hx]+\ih\int dt\hj(t)\hx(t)},
\ee
contains the action \eq{ctpaction}. The propagator $\hD$ is of the form
\be\label{ctpprop}
\hD(t-t')=-i\begin{pmatrix}\la T[x(t)x(t')]\ra&\la x(t')x(t)\ra\cr\la x(t)x(t')\ra&-\la T[x(t)x(t')]\ra^*\end{pmatrix}
\ee
and one encounters the same equations of motion and Green functions in harmonic classical and quantum systems.

When the density matrix is sought then the trace is omitted in the generator functional \eq{ctpgf} and the corresponding path integral expression contains open time paths (OTP) \cite{feynmanv,nv}.

\subsection{Decoherence}
Decoherence, the suppression of the off-diagonal elements of the density matrix in a given base, refers to the state of the system at a given time and the OTP path integral offers a dynamical view of the establishment of this decohered state. In fact, parametrize the OTP trajectories by means of the physical and the decoherence trajectories, $x(t)$ and $y(t)$, respectively, as $x^\pm(t)=x(t)\pm y(t)/2$ and write the open time path path integral of a harmonic system as
\be\label{otppi}
\int D[\hx]e^{\frac{i}{2\hbar}\hx\hat K\hx}
=\int D[x]D[y]e^{\frac{i}{2\hbar}(xK^ay+yK^rx+\hf yK^ny)-\frac1{2\hbar}y\bar Ky}
\ee
where $\hat K$ is given by Eq. \eq{qctpa} and $K^{\stackrel{r}{a}}=K^n\pm K^f$. The block matrix $\bar K$ plays a remarkable role in this path integral:

\begin{enumerate}
\item {\em Decoherence:} The imaginary part of the CTP action governs the suppression of the off diagonal elements of the density matrix. Therefore $\bar K$ is responsible of the gradual building up the decoherence in the final state, namely the consistency of trajectories \cite{griffith,omnes,gellmann}. 

\item {\em Irreversibility:} $\bar K(\omega)$, being the imaginary part of the inverse Feynman propagator, is actually the inverse life-time of the mode of frequency $\omega$. Due to the CTP consistency equation \eq{ctpcons} decoherence (suppression of paths with $y\ne0$ in Eq. \eq{otppi}) and irreversibility (normal modes of finite life-time) have the same dynamical origin and the thermodynamical and quantum time arrows are identical.

\item {\em Quantum and classical fluctuations:} Classical fluctuations are encoded in the common initial conditions of the CTP doublet trajectory $\hx$, by means of $x(0)$ and $\dot x(0)$ and are independent of $\bar K$. Quantum fluctuations emerge during the time evolution as the difference between the two members of the CTP doublet, $x^+(t)-x^-(t)$, and are governed by $\bar K$. 
\end{enumerate}

\subsection{Soft quantum-classical transition}
The CTP formalism suggests an unusual property of the quantum-classical transition. The classical limit of the path integral representation of the transition amplitude,
\be\label{fpi}
\la x_f|e^{-\ih Ht_f}|x_i\ra=\int_{x(0)=x_i}^{x(t_f)=x_f}D[x]e^{\ih S[x]},
\ee
used to be identified by the limit $\hbar\to0$. The transition amplitude becomes dominated in this limit by the trajectories within a small neighborhood of the classical trajectory, $x(t)\to x_{cl}(t)$, a feature which can be summarized by saying that the trajectory is rigid, it supports only small fluctuations in this limit.

The picture of the transition, displayed by the OTP path integral for the density matrix,
\be\label{spi}
\rho(x^+,x^-)=\int_{x^\pm(t_f)=x^\pm}D[\hx]e^{\ih S[x^+]-\ih S^*[x^-]+\ih S_i[\hx]}
\ee
where the influence functional $S_i[\hx]$ represents the environment is just the opposite. In fact, by assuming that the influence functional induces decoherence in the coordinate representation, $x^+(t)\sim x^-(t)$ we can approximate the diagonal elements of the density matrix as a path integral over the common trajectory $x(t)=x^\pm(t)$,
\be\label{clctpint}
\rho(x_f,x_f)\sim\int_{x(t_f)=x_f}D[x]e^{\ih S_i[x,x]}.
\ee
We have weak dependence on the trajectory in the exponent of integrand, $S_i[\hx]=\ord{g}$ where $g$ is the system-environment coupling strength. A wide range of trajectories contribute to the transition probabilities, in other words the classical limit is soft. One encounters here a delicate situation, known from Statistical Physics, where the system-environment interactions must be strong enough to drive the system to equilibrium but weak enough to ignore the environment in equilibrium. In our case the dependence of the influence functional $S_i[\hx]$ on $\hx$ is supposed to be strong enough to decohere the system but the numerical value, $S_i[\hx]=\ord{g}$ is small enough for weakly coupled system-environment to observe large fluctuations in the path integral \eq{clctpint}.

I believe that the resolution of this apparent conflict is to recall that the classical limit can not be reached by means of transition amplitudes between pure states, \eq{fpi}, rather we need the reduced density matrix, \eq{spi}. The formal limit $\hbar\to0$ of \eq{fpi} is not the classical one. Another indication of the softness of the classical limit comes from operator formalism: The classical limit is supposed to be reached when the excitation level density of the whole system is high. We need then small energy to orthogonalize the whole system by a jump from one stationary state to another. This is the hallmark of softness.

The softness of the trajectories in the path integral \eq{clctpint} indicates that the quantum-classical transition is in the strong coupling regime. But one must be able to recover the traditional weakly coupled classical systems from quantum mechanics. For this to happen the trajectories should become stiff as we enter into the classical domain. Such a strongly coupled crossover beside a weakly coupled regime may hide a change of degrees of freedom. In fact, this is just what happens when the asymptotically free quark-gluon description of the strong interaction changes into the hadronic scheme around a characteristic length scale, the nucleon size. This possibility raises the question whether the degrees of freedom of the classical and quantum domains are the same.

\section{Summary}\label{summs}
The emergence of the thermodynamical and quantum time arrow was demonstrated in the framework of a simple harmonic system. The time arrow is induced by the environment in the same manner as the symmetry is broken spontaneously by infinitesimal explicit symmetry breaking. To generalize this mechanism for realistic many-body systems one needs the CTP formalism which yields a systematic treatment of initial condition problems both on classical and on quantum level. The harmonic oscillators, used in this work become quasi particles when interactive quantum field theoretical models are considered, an unavoidable step in view of the large number of particles, involved in macroscopic bodies.

The structure of the CTP propagator assures the identity of the thermodynamical and quantum time arrow, an issue similar to the conjecture that statistical physics exists in the quantum domain only \cite{gemmer}.

It was argued that the quantum-classical crossover is strongly coupled. This possibility represents a wonderful challenge to revisit the correspondence principle and to derive classical physics in a quantum field theoretical framework.

\end{document}